\documentclass [12pt]{article}
\usepackage {graphicx}
\usepackage {amsmath,amssymb}

\begin{document}
\begin{center}
\subsection*{INSULATOR-METAL TRANSITION IN A CONSERVATIVE SYSTEM: AN EVIDENCE FOR
MOBILITY COALESCENCE IN ISLAND SILVER FILMS}

Manjunatha Pattabi

Materials Science Department

Mangalore University

Mangalagangotri 574 199 INDIA

\end{center}

\subsection*{Abstract:}

Aging, which manifests itself as an irreversible increase in electrical
resistance, in island metal films is of considerable interest from both
academic as well as applications point of view. Aging is attributed to
various causes, oxidation of islands and mobility of islands followed by
coalescence (mobility coalescence) being the main contenders. The effect of
parameters like substrate temperature, substrate cleaning, residual gases in
the vacuum chamber, ultrasonic vibration of the substrate suggest that the
mobility coalescence is responsible for the aging in island metal films.
Electron microscopy studies show evidence for mobility of islands at high
substrate temperatures. The comparison of aging data of island silver films
deposited on glass substrates in ultra high vacuum and high vacuum suggests
that the oxidation of islands, as being responsible for aging in these
films, can be ruled out. Further, under certain conditions of deposition,
island silver films exhibit a dramatic and drastic fall in electrical
resistance, marking the insulator-metal transition. This interesting
transition observed in a conservative system-after the stoppage of
deposition of the film- is a clear evidence for mobility coalescence of
islands even at room temperature. The sudden fall in resistance is preceded
by fluctuations in resistance with time and the fluctuations are attributed
to the making and breaking of the percolation path in the film.

\vskip 1 cm
 \noindent
 e-mail: manjupattabi@hotmail.com

\subsection*{Introduction:}

Thin film growth on any substrates occurs in one of the two broad categories
of growth modes namely Layer by Layer growth mode or Stranski- Krastanov
growth mode and island growth mode or Volmer-Weber growth mode. Surface and
interfacial energies determine the growth mode of a film. In general, vapor
deposition of metal films on glass substrates grow by island mode. In the
island mode, vapor atoms impinging on the substrate surface get adsorbed and
are known as adatoms. These adatoms migrate on the substrate surface to form
nuclei. When two nuclei touch each other they coalesce to form a larger
cluster. As the deposition continues, at a particular surface coverage, a
Large Scale Coalescence (LSC) takes place forming a network structure
leaving holes and channels in between. Secondary nucleation in the holes and
channels fill up to form a continuous film.

Therefore, by limiting the growth of a film to nucleation stage or by
avoiding excessive coalescence island films or discontinuous films,
consisting an array of discrete islands with statistical distribution of
island radii and separations, can be obtained. Although, such films have
many attractive properties, they cannot by exploited yet for device
applications due to their inherent temporal instability or aging even in
vacuum. Aging manifests as an irreversible increase in their DC electrical
resistance.

\textbf{Aging Studies:}

The expression for electrical resistance of an island film is exponentially
dependent on the average inter-island spacing [1, 2] and aging is attributed
to the increase in average island spacing following the stoppage of
deposition. The increase in the average island spacing can come about by
island shape changes [3], oxidation of islands [4] and mobility of islands
followed by coalescence [5].

Oxidation of islands model and mobility coalescence model are the main
contenders in explaining the aging process in island metal films. Studies on
the effect of residual gases on the aging of island silver and copper films
through the quantification of the aging process indicated that oxidation
might not be responsible for aging in these films [6, 7]. The substrate
temperature effect too supported this view [8].

As the deposited material in the film is increased, the average island size
increases and average island spacing decreases resulting in lower resistance
of the film. If mobility coalescence is operative, a reduced aging rate is
expected for higher thickness or lower resistance films, as mobility is size
dependent, larger islands being less mobile. If the deposition continues, a
LSC would occur and a very small aging rate is expected due to the
incorporation of small islands to the network structure.

Table 1 shows typical coalescence rates, defined as the change in tunneling
length per minute calculated over fixed time interval [9], for silver films
under different conditions of ultrasonic vibrations of the substrate. Here,
NV refers to no vibration of the substrate either during deposition or
during aging. VD is the condition where in substrate is vibrated only during
deposition. VA refers to the vibration only during aging [10]. It is clear
from the table that the larger islands formed due to vibration during
deposition show lower coalescence rate as compared to NV condition. The
films formed under similar conditions show higher coalescence rate when the
films are subjected to vibration during aging (NV and VA). Further, lower
initial resistance (resistance immediately after the deposition is stopped)
films show lower coalescence rate as they contain larger islands. These
observations strongly support mobility coalescence model.

\subsection*{Table 1 : Coalescence rates for silver island films under different
conditions of substrate vibrations}

\begin{table}[htbp]
\begin{tabular}
{|p{90pt}|p{90pt}|p{80pt}|p{80pt}|} \hline Initial Resistance&
\multicolumn{3}{|p{250pt}|}{Coalescence Rates ({\AA}/min)}  \\
\hline
&
NV&
VD&
VA \\
\hline 2 M$\Omega / \square $ & 0.0656& 0.0345&
0.0889 \\
\hline 10 M$\Omega / \square $ & 0.1284& 0.08321&
0.1566 \\
\hline 20 M$\Omega / \square $ &
 0.2209& 0.1575&
0.3179 \\
\hline
\end{tabular}
\label{tab1}
\end{table}

The silver island films on glass substrates studied under a UHV of 2x10$^{ -
8}$ Torr with an oxygen partial pressure of the order of 10$^{ - 11}$ Torr
too show considerable aging for extended periods. Further, coalescence rates
under UHV and HV of 2x10$^{ - 6}$ Torr -- HV obtained using an oil diffusion
pump- are nearly the same [11]. These results further strengthen the
mobility coalescence model.

\subsection*{Insulator-Metal Transition:}

It is well known that after the occurrence of LSC very little
material is required to be deposited to form a semicontinuous film
with a connecting path between the electrodes being established.
This is the discontinuous -- semicontinuous transition often
termed as insulator -- metal transition in the case of granular
metal films. Island silver films deposited on glass substrates
with an initial resistance of 2 M$\Omega / \square$, at a
deposition rate of 0.2 {\AA}/s exhibits a very interesting
behavior and is shown in Fig. 1. The film shows a small aging rate
for a few minutes of aging. Then the resistance falls drastically,
after which the resistance remains almost steady. The aging is
characteristic of island or discontinuous films while a steady or
a steady decrease in resistance with time is the property of a
continuous or a semicontinuous film. The behavior exhibited by
this film clearly indicates an insulator -- metal transition. The
small aging rate observed before the transition indicates that the
LSC stage has already been crossed for this film. At a very low
deposition rate of 0.2 {\AA}/s agglomeration would be much less.
Therefore, one can expect LSC to occur at an earlier stage itself
as the degree of agglomeration determines the thickness at which
film tends to become continuous [12]. However, insulator -- metal
transition occurring after the stoppage of deposition is a
fascinating new result.

\begin{figure}[htbp]
\centerline{\includegraphics[width=5.32in]{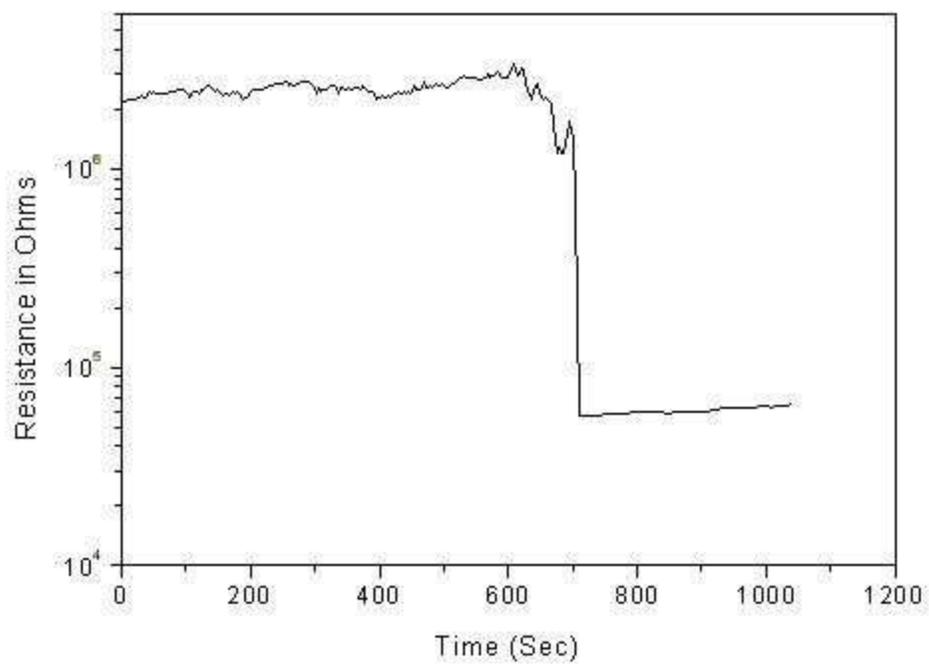}}
\label{Fig1} \caption{The variation of resistance with time after
the stoppage of deposition for a Silver film of thickness $\sim
$100 {\AA}}
\end{figure}

The insulator -- metal transition in a conservative system may be
explained as follows. At very low deposition rates, not only LSC
occurs at an early stage of deposition, but, due to reduced
agglomeration, very little material is required to connect the large
network to establish a percolation path between the electrodes. The
incorporation of secondary nuclei into the network structure due to
their mobility may result in a percolation path leading to an
insulator metal transition. This cannot be expected in a more
agglomerated film deposited at higher deposition rates as more
material is required to establish a percolation path and just the
incorporation of secondary nuclei would not be sufficient for this
purpose. As a consequence, insulator -- metal transition after the
stoppage of deposition is not observed for films deposited at higher
rates (eg. Ref. 10)

The other interesting feature is that the transition is not a gradual
decrease of resistance but a sudden one and is preceded by fluctuations in
resistance with time. These fluctuations in resistance can, in principle, be
due to the dynamic fluctuations in the average island spacing. But, such
dynamic fluctuations in average island spacing due to the mobility of
islands should always be present, not only in the vicinity of insulator -
metal transition. Under the conditions that are not leading to such a
transition, the variation of resistance with time were always found to be
smooth. On the other hand, these fluctuations can be due to the making and
breaking of the circuit or the conducting paths between the networks. The
decrease in resistance is due to the filling up of the gaps between the
networks by the secondary nuclei. The gaps filled by these secondary nuclei
forms a weak electrical link between two large networks. The typical size of
secondary nuclei can be from a few to a few tens of nanometers. The current
used by the resistance-measuring instrument used in the present studies
(Keithley Eectrometer model 617) is only of the order of microamperes. But
still, the current density flowing through these weak links would be very
large ($\sim $10$^{6 }$A/cm $^{2}$ for the 10 nm cluster) and can cause the
catastrophic destruction of these weak links. This would give rise to the
breaking of the conduction paths established, increasing the film
resistance. At the onset of insulator - metal transition, a large number of
parallel conducting paths between the electrodes are established and
therefore, the current density gets distributed, avoiding the destruction of
the weak links. Therefore, the film resistance remains steady after the
transition.

\subsection*{Conclusions:}

The conclusions that can be drawn from the aging studies of silver island
films can be summarized as follows:

1. The effect of various parameters on the aging rates of Silver Island
films suggest that the mobility of islands followed by coalescence is
responsible for aging in these films.

2. An interesting insulator - metal transition is observed under certain
conditions in silver films, long after the stoppage of deposition. This is
attributed to the establishment of metallic conduction paths due to the
incorporation of secondary nuclei to the network structure. The transition
is preceded by large fluctuations in resistance.

3. The fluctuations in film resistance is attributed to the formation of
conduction paths and their breakage due to the passage of high current
density at the weak links in the path.

4. The insulator-metal transition observed in a conservative system is a
clear evidence for the mobility of islands.

\subsection*{Acknowledgements: }

The author would like to thank Dr.B.A.Dasannacharya, Director, IUC-DAEF, for
useful discussions and for his constant encouragement. Thanks are also due
to Prof.Ajay Gupta, Director, IUC-DAEF, Indore Centre, and Dr.S.M. Chaudhuri
for their collaboration in the insulator-metal transition work.

\subsection*{References:}
\begin{enumerate}

    \item C.A. Naugebauer and M.B.Webb, J.Appl.Phys. \textbf{33} (1962) 74

    \item R.M.Hill, Proc.R.Soc. London, Ser.A.\textbf{309} (1969) 377

    \item M.Nishiura and A.Kinbara, Thin Solid Films, \textbf{24} (1974) 75

    \item F.P.Fehlner, J.Appl.Phys.\textbf{ 38} (1967) 2223

    \item J.Skofronick and W.B.Phillips, J.Appl.Phys.\textbf{ 38} (1967) 4791

    \item V.Damodara Das and M.S.Murali Sastry, J.Appl.Phys.\textbf{ 59} (1986) 3184

    \item V.Damodara Das and M.S.Murali Sastry, Phys.Rev.B.\textbf{ 33} (1986) 6612

    \item M.Pattabi, M.S.Murali Sastry and V.Sivaramakrishnan, J.Appl.Phys.\textbf{
63} (1988) 893

    \item M.Pattabi and M.S.Murali Sastry, Thin Solid Films, \textbf{159} (1988) L61

    \item M. Pattabi, J.Uchil and K.Mohan Rao, Thin Solid Films, \textbf{305 }(1997)
196

    \item M.Pattabi et al, Thin Solid Films, \textbf{322} (1998) 340

    \item K..Chopra, Thin Film Phenomena, Robert E Krieger Publishing, New York, 1979,
pp172-173

\end{enumerate}

\end{document}